\documentclass[12pt]{article}
\usepackage{amsmath, amsthm, amsfonts}
\usepackage{graphicx}            
\usepackage{epsfig}

\newtheorem{thm}{Theorem}
\newtheorem{dfn}[thm]{Definition}
\newtheorem{lem}[thm]{Lemma}
\newtheorem{prp}[thm]{Proposition}

\newtheorem{hyp}{Hypothesis}
\theoremstyle{remark}
\newtheorem{rem}{Remark}

\newcommand{\paren}[1]{\left(#1\right)}
\newcommand{\norm}[1]{\left\Vert#1\right\Vert}
\newcommand{\abs}[1]{\left|#1\right|}
\newcommand{\set}[1]{\left\{#1\right\}}

\def\fa{\; \forall \;}

\def\RR{\mathbb{R}}
\def\to{\rightarrow}
\def\L1{L^1[0,T]}
\def\one{1}

\def\e{\epsilon}

\def\l{\lambda}

\title{Asymptotic behavior of solutions to the generalized
  Becker-D\"oring equations for general initial data}

\author{Jos\'e Alfredo Ca\~nizo Rinc\'on\\
  \small Dept. Matem\'atica Aplicada\\
  \small Universidad de Granada\\
  \small E18071 Granada\\
  \small Spain
}

\date{10 July 2005}

\begin{document}
\maketitle

\abstract{We prove the following asymptotic behavior for solutions to
  the generalized Becker-D\"oring system for general initial data:
  under a detailed balance assumption and in situations where density
  is conserved in time, there is a critical density $\rho_s$ such that
  solutions with an initial density $\rho_0 \leq \rho_s$ converge
  strongly to the equilibrium with density $\rho_0$, and solutions
  with initial density $\rho_0 > \rho_s$ converge (in a weak sense) to
  the equilibrium with density $\rho_s$. This extends the previous
  knowledge that this behavior happens under more restrictive
  conditions on the initial data. The main tool is a new estimate on
  the tail of solutions with density below the critical density.}

\section{Introduction}

Coagulation-fragmentation equations are useful as models that describe
the dynamics of many physical phenomena in which a large number of
particles or units can stick together to form groups of particles, or
clusters. A first version of them was initially proposed by Becker and
D\"oring \cite{BD35}, and a variant by Penrose and Lebowitz
\cite{PL79}; these relatively simple models take into account only
processes in which a cluster gains or loses one particle, and describe
only the concentration of clusters of a given size at a certain
moment, omitting also a description of their spatial distribution.
Since then a number of generalizations have been studied which also
allow reactions between clusters of more than one particle, the main
examples of this being the discrete coagulation-fragmentation
equations (see for example \cite{BC90,C92,CdC94}), their continuous
version \cite{L00,L02,S17,S89,S90,EMP02,ELMP,MR03} and the respective
versions including a spatial description by means of diffusion
\cite{LM02,LM02d}. A recent review can be found in \cite{LM04}.

The generalized Becker-D\"oring equations are an intermediate step
between the Becker-D\"oring system and the full discrete
coagulation-fragmen\-ta\-tion equations in which we allow reactions
between clusters of at most a given finite size $N$ and other
clusters. The system of equations is the following:
\begin{align}\label{eq:gBD}
    & \dot{c}_1 = -\sum_{k=1}^{\infty} W_{1,k} & \\
    & \dot{c}_j = \frac{1}{2} \sum_{k=1}^{j-1} W_{j-k,k} -
                 \sum_{k=1}^{\infty} W_{j,k}, \qquad   & 2\leq j \leq N \notag \\
    & \dot{c}_j = \frac{1}{2} \sum_{k=1}^{j-1} W_{j-k,k} -
                 \sum_{k=1}^{N} W_{j,k}, \qquad   & N+1 \leq j \leq 2N \notag \\
    & \dot{c}_j = \sum_{k=1}^{N} W_{j-k,k} -
                 \sum_{k=1}^{N} W_{j,k}, \qquad   & j \geq 2N+1
                 \notag
\end{align}
Here the unknowns are $c_j = c_j(t)$ for $j = 1,\dots$, positive
functions depending on the time $t$ which are intended to represent
the density of clusters of size $j$ (those formed by $j$ elementary
particles). The quantities $W_{jk}$, which depend on the $c_j$, are
given by
\[
  W_{jk} := a_{jk} c_j c_k - b_{jk} c_{j+k} \quad (j,k \geq 1),
\]
where the numbers $a_{jk}, b_{jk}$ for $j,k \geq 1$ with $\min\{j,k\}
\leq N$ are the coagulation and fragmentation coefficients,
respectively, which are symmetric in $j,k$. As can be seen, this
system is a particular case of the coagulation-frag\-men\-ta\-tion
equations when $a_{jk}=b_{jk}=0$ if $\min\{j,k\}>N$.

The study of the long-time behavior of solutions to these equations is
expected to be a model of physical processes such as phase transition.
Call $\sum_{j=1}^\infty j c_j$ the \emph{density} of a solution
$\{c_j\}_{j \geq 1}$. For the Becker-D\"oring equations it was proved
in \cite{BCP86} and \cite{BC88} that, under certain general conditions
which include a detailed balance (see below), there is a critical
density $\rho_s \in [0,\infty]$ such that any solution that initially
has density $\rho_0 \leq \rho_s$ ($\rho_0 < \infty$ if $\rho_s =
\infty$) will converge for large times, in a certain strong sense, to
an equilibrium solution with density $\rho_0$, while any solution with
density above $\rho_s$ will converge (in a weak sense) to the only
equilibrium with density $\rho_s$. The rate of convergence to
equilibrium was studied in \cite{JN03}. The mentioned weak convergence
can then be interpreted as a phase transition in the physical process
modelled by the equation (see below for a precise statement). It is an
interesting problem to extend this result to more general models; this
has been done for the generalized Becker-D\"oring equations in
\cite{CdC94} under some conditions on the decay of the initial data
and in \cite{dC98} for suitably small initial data. The aim of this
paper is to prove that this result about the generalized
Becker-D\"oring system is true for general initial data. The
corresponding result is expected to hold for the full
coagulation-fragmentation equations, but finding
a proof of this is still an open problem.\\
\textbf{Acknowledgements.}These results were obtained under the
supervision and help of St\'ephane Mischler. I wish to thank him for
his explanations and suggestions. I would also like to thank the
anonymous referees for their valuable corrections and comments. The
author was supported by FPU grant AP2001-3940 and by EU financed
network no. HPRN-CT-2002-00282.

\section{Statement of the main result}
\label{sec:statement}

Let us recall some usual definitions and notation from previous works
on the coagulation-fragmentation equations. We will make use of the
vector space
\begin{equation*}
  X := \set{ \{c_j\}_{j \geq 1} \,\Big\vert\,
    \sum_{j=1}^\infty j \, \abs{c_j} < \infty }
\end{equation*}
with norm
\begin{equation*}
  \quad \norm{c} := \sum_{j=1}^\infty j \, \abs{c_j}
  \quad \fa c = \{c_j\}_{j \geq 1} \in X.
\end{equation*}
The space $X$ is clearly a Banach space (actually, this space is
isometric to the space of absolutely summable sequences under the map
$\{c_j\} \mapsto \{j\,c_j\}$). In it we will make use of the notion of
convergence associated to the norm $\norm{\cdot}$, which we will call
``strong convergence'' following common usage.  We will also say that
a sequence $\{c^i\}_{i \geq 1}$ of elements of $X$ converges weak-$*$
to an element $c \in X$, and will denote it by $c^i
\overset{*}{\rightharpoonup} c$, if
\begin{enumerate}
\item there exists $M \geq 0$ such that $\norm{c^i} \leq M$ for all $i
  \geq 1$ and
\item $c^i_j \to c_j$ when $i \to \infty$, for all $j \geq 1$ (where
  $c^i = \{c^i_j\}_{j \geq 1}$ and $c = \{c_j\}_{j \geq 1}$).
\end{enumerate}
This is just the usual weak-$*$ convergence in the space $X$ when it
is regarded as the dual space of the space of sequences $\{c_k\}_{k
  \geq 1}$ such that $\lim_{k \to \infty} k^{-1}\, c_k = 0$, with norm
given by $\norm{\{c_k\}} := \max \set{ k^{-1}\, \abs{c_k} \mid k \geq
  1 }$ (see \cite{BCP86}, p. 672). We also cite a result from
\cite{BCP86}:
\begin{lem}[\cite{BCP86}, Lemma 3.3]
  \label{lem:weak-strong-convergence}
  If $\{c^n\}$ is a sequence in $X$ such that $c^n
  \overset{*}{\rightharpoonup} c \in X$ and $\norm{c^n} \to \norm{c}$,
  then $c^n \to c$ strongly in $X$.
\end{lem}

The subset of $X$ formed by the sequences of nonnegative terms will be
referred to as $X^+$:
\begin{equation*}
  X^+ := \set{ \{c_j\}_{j \geq 1} \in X
    \mid c_j \geq 0 \,\,\,\forall\, j \geq 1}.
\end{equation*}

We will ask for any solution $\{c_j(t)\}_{j \geq 1}$ to be, for each
fixed time $t$, in $X^+$; this is natural, given that densities should
be positive and that the sum $\sum_{j=1}^\infty j \, c_j(t)$
represents the total density of particles at time $t$ (or total mass,
depending on the interpretation given to the $c_j$'s). More precisely,
we will use the following concept of solution from \cite{BC90},
section 2:
\begin{dfn}
  A solution on the interval $[0,T[$ (for a given $T > 0$ or
  $T=\infty$) of (\ref{eq:gBD}) is a function $c : [0,T[ \to X^+$ such
  that, if we put $c(t) = \{c_j(t)\}_{j \geq 1}$ for $t \in [0,T[$,
  \begin{enumerate}
  \item $c_j : [0,T[ \to \RR$ is absolutely continuous for all $j \geq
    1$ and $\norm{c(t)}$ is bounded on $[0,T[$,
  \item for all $j = 1,2,\dots$, the sums $\sum_{k=1}^\infty a_{j,k} c_k(t)$
    and $\sum_{k=1}^\infty b_{j,k} c_{j+k}(t)$ are finite for almost all $t \in [0,T[$,
  \item and equations (\ref{eq:gBD}) hold for almost all $t \in [0,T[$.
  \end{enumerate}
\end{dfn}
\begin{rem}\label{rem:results_are_valid}
  For convenience, this definition has been slightly changed with
  respect to that in \cite{BC90}: it has been stated for the
  generalized Becker-D\"oring system instead of the full coagulation
  equations, and conditions have been phrased in different terms, but
  it can easily be checked that if the coefficients $a_{jk}, b_{jk}$
  satisfy hypothesis \ref{hyp:gBD} below then this concept of solution
  is equivalent to that in \cite{BC90}. Hence, results from
  \cite{BC90} are also applicable in our case, a fact that we will use
  later.
\end{rem}

As we do not know of a uniqueness result that can be applied under the
above hypotheses we need to define a concept of admissibility to
precise which solutions our result applies to. In \cite{CdC94} this is
done by choosing solutions which are limits of solutions to the finite
set of equations obtained by truncating system (\ref{eq:gBD}). We will
call these solutions \emph{Carr--da Costa admissible}. Here we will
define a slight modification of this concept: an admissible solution
will be one which is the limit of Carr--da Costa admissible solutions
with truncated initial data. The concept must of course be the same
under any set of conditions that ensure uniqueness, but we have not
found a sufficiently general uniqueness result and thus the following
will be needed:
\begin{dfn}\label{dfn:admissible}
  Take $T>0$ or $T = +\infty$. An admissible solution of the
  generalized Becker-D\"oring equations (\ref{eq:gBD}) on $[0, T[$
  with initial data $c^0 = \{c^0_j\}_{j \geq 1} \in X^+$ is a solution
  $c$ which is a limit in $L^\infty_{\text{loc}}([0,T[, X)$ of
  Carr--da Costa admissible solutions $c_n = \{c_n^j\}_{j \geq 1}$ of
  (\ref{eq:gBD}) with truncated initial data $c^{0,n}$ given by
  \begin{align*}
    &c^{0,n}_j := c^0_j \text{ for } j \leq n\\
    &c^{0,n}_j := 0 \text{ for } j > n.
  \end{align*}
\end{dfn}
\begin{rem}\label{rem:uniform_convergence}
  The above convergence is uniform in compact subsets of $[0,T[$, in
  the sense of the norm $\norm{\cdot}$ in $X$; in particular, the
  functions $c^n_j$ in the definition converge uniformly when $n \to
  \infty$ in compact subsets of $[0,T[$ to $c_j$.
\end{rem}

Below we state the conditions on the coefficients under which we will
prove our result. Though in the equations only the coefficients
$a_{jk}, b_{jk}$ with $\min\{j,k\} \leq N$ appear, for convenience we
will use coefficients $a_{jk}, b_{jk}$ defined for all $j,k \geq 1$
and simply set $a_{jk}=b_{jk}=0$ if $\min\{j,k\}>N$. Thus we have
hypothesis \ref{hyp:gBD}:

\begin{hyp}[Generalized Becker-D\"oring] \label{hyp:gBD} There
exists an $N \geq 2$ such that $a_{jk}=b_{jk}=0$ if
$\min\{j,k\}>N$, and $a_{jk}, b_{jk}>0$ otherwise.
\end{hyp}

Detailed balance is a physical assumption also used, for example, in
\cite{BCP86, CdC94}, which expresses the principle of microscopic
reversibility from chemical kinetics; essentially, it states that
equilibria of a certain form exist (see theorem \ref{thm:equilibria}
below):
\begin{hyp}[Detailed Balance] \label{hyp:db}
  There exists a positive sequence $\{Q_j\}_{j \geq 1}$ with $Q_1=1$
  such that for all $j,k \geq 1$,
  \begin{equation}
    a_{jk} Q_j Q_k = b_{jk} Q_{j+k}.
  \end{equation}
\end{hyp}

A certain bound on the growth rate of coefficients is known to be
necessary to ensure the existence of density-conserving solutions
\cite{BC90,ELMP} (in other situations density is only conserved for a
finite time after which density decreases, a phenomenon known as
gelation); for our main result to be true (theorem \ref{thm:final}) it
is evidently necessary that density is conserved, so we impose a
condition ensuring this.
\begin{hyp}[Growth of coefficients]\label{hyp:growth}
  For some constants $K>0$ and $0\leq \alpha <1$,
  \begin{align*}
    & a_{jk} \leq K (j^\alpha + k^\alpha), \\
    & b_{jk} \leq K (j^\alpha + k^\alpha).
  \end{align*}
\end{hyp}

In the next hypothesis, (\ref{eq:Qjk1}) is a physical condition that
asserts that any cluster has a lower free energy than its pieces taken
separately (see \cite{CdC94}, Remark 5.1); (\ref{eq:Qjk2}) will be
seen to imply the existence of a critical density $\rho_s$ (the
relationship between the following $z_s$ and this critical density is
given below in \ref{thm:equilibria}):
\begin{hyp}\label{hyp:Qjk}
  The sequence $Q_j$ satisfies:
  \begin{align}
    \label{eq:Qjk1}
    & \log Q_j + \log Q_k \leq \log Q_{j+k}
       \quad \text{for all } j,k \geq 1, \\
    \label{eq:Qjk2}
    & 0 < \lim_{j\to\infty} \frac{Q_j}{Q_{j+1}} := z_s < \infty.
  \end{align}
\end{hyp}
\begin{rem}\label{rem:limitsQj}
  This implies that $\lim_{j\to\infty} \frac{Q_j}{Q_{j+m}} = z_s^m$
  for $m \geq 1$ and that $\lim_{j\to\infty} Q_j^{1/j} =
  \frac{1}{z_s}$.
\end{rem}


We also need to assume, as new hypotheses, a certain regularity
of the coefficients:
\begin{hyp}\label{hyp:ajk_regular}
  For $j,m = 1,\dots,N$,
  \[
    \frac{a_{jk}}{a_{j,k+m}} \to 1 \quad \text{when } k \to \infty
  \]
\end{hyp}
\begin{hyp}\label{hyp:ajk_regular2}
  For some constant $K_a$, $j,m = 1,\dots,N$ and $k \geq 1$,
  \[
  \abs{ a_{jk} - a_{j,k+m} } \leq K_a.
  \]
\end{hyp}
Observe that hypotheses \ref{hyp:ajk_regular} and
\ref{hyp:ajk_regular2} are independent;
for example, for $j=1,\dots,N$ and $k \geq 1$, $a_{jk} = \exp(-j-k)$
satisfies the second one but not the first; and $a_{jk} = [ \log (j+k)
] \sqrt{j+k}$ (with $[x]$ being the integer part of $x$) satisfies the
first but not the second.

\begin{rem}
  The kind of coefficients allowed by the previous hypotheses are, for
  example, $a_{jk} \leq C(j^\alpha + k^\alpha)$ for $j=1,\dots,N$ and
  $k \geq 1$, sufficiently regular to fulfill hypotheses
  \ref{hyp:ajk_regular} and \ref{hyp:ajk_regular2}, and $b_{jk}$ given
  by hypothesis \ref{hyp:db} with any choice of $Q_j$ satisfying
  (\ref{eq:Qjk1}) and (\ref{eq:Qjk2}). Note that (\ref{eq:Qjk1})
  implies that $b_{jk} \leq a_{jk}$, so $b_{jk} \leq C (j^\alpha +
  k^\alpha)$ also. For a concrete example, pick $C_1, C_2 > 0$ and
  $\alpha, \delta \in [0,1[$ and define the following coefficients for
  $\min\{j,k\} \leq N$:
  \begin{align*}
    &a_{jk} := C_1 (j^\alpha + k^\alpha)\\
    &b_{jk} := C_1 (j^\alpha + k^\alpha)
    \exp \paren{C_2 \paren{(j+k)^\delta - j^\delta - k^\delta}}.
  \end{align*}
  The coefficients are taken to be zero when $\min\{j,k\} > N$. These
  correspond to $Q_j = \exp \paren{C_2(j-j^\delta)}$ and have $z_s =
  e^{-C_2}$.
\end{rem}

We borrow known existence results for the kind of admissible solutions
of definition \ref{dfn:admissible} from \cite{BC90}:
\begin{thm}[\cite{BC90}, Theorems 2.4, 3.6 and 5.4]
  \label{thm:existence}
  Assume hypotheses \ref{hyp:gBD} and \ref{hyp:growth}, and take $c^0
  \in X^+$. Then there exists an admissible solution $c$ to
  (\ref{eq:gBD}) on $[0, +\infty[$ with $c(0) = c^0$. Furthermore,
  under hypothesis \ref{hyp:gBD} all solutions to (\ref{eq:gBD}) are
  density-conserving.
\end{thm}
\begin{rem}
  Theorem 2.4 in \cite{BC90} gives the existence of a solution (in
  fact, a Carr--da Costa admissible solution by the method of
  construction).  Theorem 3.6 from \cite{BC90} proves this solution
  conserves density.  Finally, Theorem 5.4 in the same paper gives the
  existence of a solution that can be obtained as the uniform limit in
  compact sets of $[0,T[$ of Carr--da Costa admissible solutions with
  truncated initial data, thus giving the existence of an admissible
  solution in the sense used here.
\end{rem}

\begin{lem}
  \label{lem:moments}
  Assume hypotheses \ref{hyp:gBD} and \ref{hyp:growth}. Take $\mu > 1$
  and suppose that $c = \{c_j\}_{j \geq 1}$ is an admissible solution
  to (\ref{eq:gBD}) on $[0, T[$ for some $T > 0$ with initial data
  $c(0) = c^0$ such that $\sum_{j=1}^\infty j^\mu c^0_j < +\infty$.
  Then $\sum_{j=1}^\infty j^\mu c_j(t)$ is finite for all $0 \leq t <
  T$.
\end{lem}

\begin{proof}
  This is just Theorem 3.3 in \cite{CdC94}, stated for admissible
  solutions in the sense we use here. As Carr--da Costa admissible
  solutions satisfy the estimate given in the proof of the above
  theorem in \cite{CdC94} (which depends only on $\sum_{j=1}^\infty
  j^\mu c^0_j$), we can pass to the limit and thus prove that our
  admissible solutions also satisfy it.
\end{proof}

Hypotheses \ref{hyp:gBD}--\ref{hyp:Qjk} imply those of Theorems 5.1
and 5.2 in \cite{CdC94}: (1.7) and (\textbf{H2}) in \cite{CdC94} are
always fulfilled if we assume hypothesis \ref{hyp:gBD}; (\textbf{H1})
is our \ref{hyp:growth} and (\textbf{H3}), (\textbf{H4}) from
\cite{CdC94} are contained in hypotheses \ref{hyp:gBD} and
\ref{hyp:Qjk} here, respectively. This enables us to use these
theorems here (recall remark \ref{rem:results_are_valid}); we will
need the following one about the equilibrium solutions of
(\ref{eq:gBD}).
\begin{dfn}
  An equilibrium of (\ref{eq:gBD}) is a solution of (\ref{eq:gBD})
  that does not depend on time.  The density of an equilibrium $c$ is
  the norm of $c$ in $X$, $\sum_{j=1}^\infty j \, c_j$.
\end{dfn}
\begin{dfn}
  The critical density $\rho_s$ is defined to be
  \begin{equation*}
    \rho_s := \sum_{j=1}^\infty Q_j z_s^j,
    \quad \quad (0 < \rho_s \leq \infty).
  \end{equation*}
\end{dfn}
\begin{thm}[\cite{CdC94}, Theorem 5.2]
  \label{thm:equilibria}
  Assume hypotheses \ref{hyp:gBD}--\ref{hyp:Qjk}.
  \begin{enumerate}
  \item For $0 \leq \rho \leq \rho_s$ (and also $\rho < +\infty$ if
    $\rho_s = +\infty$), there exists exactly one equilibrium
    $\{c_j^\rho\}$ of (\ref{eq:gBD}) with density $\rho$, which is
    given by
    \begin{equation*}
      c_j^\rho = Q_j z^j \quad \forall \, j \geq 1,
    \end{equation*}
    where $z$ is the only positive number such that $\sum_{j=1}^\infty
    j Q_j z^j = \rho$.
  \item For $\rho_s < \rho < +\infty$ there is no equilibrium of
    (\ref{eq:gBD}) with density $\rho$.
  \end{enumerate}
\end{thm}
Observe that when $\rho_s$ is finite and $\{c_j^{\rho_s}\}_{j \geq 1}$
represents the critical equilibrium (the one with density $\rho_s$),
$z_s$ is the single particle density $c_1^{\rho_s}$ of this
equilibrium.

The main result in this paper is the following:
\begin{thm}\label{thm:final}
  Assume hypotheses \ref{hyp:gBD}-\ref{hyp:ajk_regular2}, and let $c =
  \{c_j\}_{j \geq 1}$ be an admissible solution of the generalized
  Becker-D\"oring equations \eqref{eq:gBD} (whose existence is given
  by theorem \ref{thm:existence}). Call $\rho_0 := \sum_{j=1}^\infty j
  c_j(0)$, the initial density.
  \begin{enumerate}
  \item If $0 \leq \rho_0 \leq \rho_s$ then $c$ converges strongly in
    $X$ to the equilibrium with density $\rho_0$.
  \item If $\rho_s < \rho_0$ then $c$ converges in the weak-$*$
    topology to the equilibrium with density $\rho_s$.
  \end{enumerate}
\end{thm}

\section{Proofs}

The following result from \cite{CdC94} already gives part of Theorem
\ref{thm:final}. Again, note that the hypotheses in \cite{CdC94} are
contained in those here:
\begin{thm}[\cite{CdC94}, Theorem 6.1]
  \label{thm:weak-star-convergence}
  Assume hypotheses \ref{hyp:gBD}-\ref{hyp:Qjk}. Let $c = \{c_j\}$ be
  a solution of (\ref{eq:gBD}) on $[0,\infty[$, and call $\rho_0 :=
  \sum_{j=1}^\infty j \, c_j$.

  Then there exists $0 \leq \rho \leq \min\set{\rho_0, \rho_s}$ such
  that $c \overset{*}{\rightharpoonup} c^\rho$, where $c^\rho$ is the
  only equilibrium of (\ref{eq:gBD}) with density $\rho$ (given by
  Theorem \ref{thm:equilibria}).
\end{thm}

With Theorem \ref{thm:weak-star-convergence}, the next result will be
enough to complete a proof of Theorem \ref{thm:final}:

\begin{thm}\label{thm:main}
  Assume hypotheses \ref{hyp:gBD}--\ref{hyp:ajk_regular2} hold.
  Suppose that $c$ is an admissible solution to the generalized
  Becker-D\"oring equations \eqref{eq:gBD} with initial data $c^0 \in
  X_+$ such that $c$ converges weak-$*$ to an equilibrium with density
  $\rho < \rho_s$. Then, $c$ converges strongly to this equilibrium
  (and in particular, $\rho$ is the density of the solution $c$, i.e.
  $\rho = \rho_0$).
\end{thm}

Hence, the aim of the rest of this section will be to prove Theorem
\ref{thm:main}. The following key result gives a bound on the
solutions that will easily imply the precompactness of the orbits,
which in turn implies Theorem \ref{thm:main}. Call, for $i \geq 1$,
\begin{equation*}
  G_i(t) \equiv \sum_{j=i}^\infty j c_j(t).
\end{equation*}

\begin{prp}\label{prp:G}
  Let $c = \{c_j\}_{j \geq 1}$ be an admissible solution of the
  generalized Becker-D\"oring equations \eqref{eq:gBD}. Assume
  hypotheses \ref{hyp:gBD}-\ref{hyp:ajk_regular2}.

  Suppose that for some $z < z_s$
  \[
    c_j(t) \leq z_j := z^j Q_j
      \quad \text{for all } j=1,\dots,N \text{ and all } t \geq 0.
  \]

  Suppose that $\{r_i\}_{i \geq 1}$ is a strictly decreasing
  sequence of positive numbers that satisfy, for some $\lambda$
  with $1 < \lambda < z_s/z$:
  \[
    \frac{r_{k-1}-r_k}{r_k-r_{k+1}} < \lambda
    \quad \text{ for all } k
  \]
  and such that $G_i(0) \leq r_i$ for all $i$.

  Then there exist a positive integer $k_0$ and a constant $C>0$ such
  that $G_i(t) \leq C r_i$ for all $i \geq k_0$ and all positive
  times.
\end{prp}


The proof of proposition \ref{prp:G}, which contains the core of the
argument, is a generalization of a method used in unpublished notes by
Ph. Lauren\c{c}ot and S. Mischler \cite{LM}. This method is inspired
by the proof of uniqueness of solutions to the Becker-D\"oring
equation in \cite{LM02e}. The use of this kind of argument can be
traced back to \cite{BC88}.

Note that the condition on $\{r_k\}$ in proposition \ref{prp:G} is not
very stringent as the following lemma states:

\begin{lem}\label{lem:findrk}
  Given $\lambda > 1$ and a positive sequence $\{g_k\}_{k \geq 1}$
  which tends to zero as $k$ tends to infinity, there exists a
  strictly decreasing positive sequence $\{r_k\}_{k \geq 1}$ which
  converges to zero, such that $g_k \leq r_k$ and
\[
  \frac{r_{k-1}-r_k}{r_k-r_{k+1}} \leq \lambda
    \quad \text{ for all } k
\]
\end{lem}

\begin{proof}
  Define
  \begin{align*}
    \bar{g}_1 &:= \sup_{j \geq 1} \{g_j\} + 1 \\
    \bar{g}_k &:= \sup_{j \geq k} \{g_j\},
       \quad \text{ for } k \geq 2 \\
    h_k &:= \bar{g}_k - \bar{g}_{k+1},
       \quad \text{ for } k \geq 1. \\
  \end{align*}
  Then $\bar{g}_k$ is decreasing, tends to zero and for all $k$ we have
  $\bar{g}_k = \sum_{j=k}^\infty h_j$. Define $s_k$ recursively as:
  \begin{align*}
    s_1 &:= h_1 \\
    s_{k+1} &:= \max \left\{ \frac{s_k}{\lambda}, h_{k+1} \right\}.
  \end{align*}
  Then $s_k>0$ for all $k$ (it is to ensure this that we added 1 to
  $\bar{g}_1$) and we can see that $\sum_{k \geq 1} s_k$ converges.
  For this, note that $s_{k+1} \leq (s_k / \lambda) + h_{k+1}$ and
  write for $m \geq 2$:
  \begin{equation*}
    \sum_{k=1}^{m+1} s_k
    = s_1 + \sum_{k=1}^{m} s_{k+1}
    \leq h_1 + \sum_{k=1}^m h_{k+1} + \frac{1}{\lambda} \sum_{k=1}^m s_k
    \leq \bar{g}_1 - \bar{g}_{m+2} +  \frac{1}{\lambda} \sum_{k=1}^m s_k,
  \end{equation*}
  so we have that
  \begin{equation*}
    \paren{ 1 - \frac{1}{\lambda} } \sum_{k=1}^m s_k \leq \bar{g}_1,
  \end{equation*}
  which proves the summability of $\{s_k\}$ since $\lambda > 1$. (I
  thank the referees for suggesting a simpler version of this proof).


  Clearly, $s_k \geq h_k$. Let us finally define
  \[
  r_k := \sum_{j=k}^\infty s_j
  \geq \sum_{j=k}^\infty h_j
  = \bar{g}_k
  \geq g_k,
  \]
  which is positive, greater than $g_k$, strictly decreasing, tends
  to zero as $k \to \infty$ and
  \[
    \frac{r_{k-1}-r_k}{r_k-r_{k+1}} = \frac{s_{k-1}}{s_k} \leq \lambda.
  \]
\end{proof}


\subsection{Proof of the proposition}

We will prove the proposition for solutions whose initial data is a
truncation at a sufficiently large finite size of $\{ c_i(0) \}_{i
  \geq 1}$, with constants $C$ and $k_0$ that do not depend on the
size of this truncation; then the proposition follows for general
initial data by a standard approximation argument using definition
\ref{dfn:admissible} of an admissible solution.

Take an $L \geq 1$ and consider a solution $\{c^L_i\}_{i \geq 1}$ with
initial data $c^L_i(0) = c_i(0)$ for $i=1,\dots,L$ and $c^L_i(0) = 0$
for $i > L$. It is again enough to prove the bound in the result up to
a finite time $T>0$, with a constant that does not depend on $T$. So
fix $T>0$, and let us find $C$ and $k_0$ (independent of $L$ and $T$)
such that
\begin{equation*}
  c^L_i(t) \leq C r_i
  \quad \text{ for all } i \geq k_0,
  \quad t \in [0,T]
  \text{ and } L \text{ sufficiently large.}
\end{equation*}
By the admissibility of $c$ we know that the functions $c_j^L$
converge uniformly in $[0,T]$ to $c_j$ as $L \to \infty$ (see remark
\ref{rem:uniform_convergence}), so the hypotheses of the proposition
imply that for sufficiently large $L$
\[
c_j^L (t) < z^j Q_j
\, \text{ for } j=1,\dots,N, \, t \in [0,T].
\]
In the following $L$ will always be large enough for this to hold
(note that the choice of $L$ depends also on $T$).

Furthermore, $G_i^L(0) \leq G_i(0) \leq r_i$ for all $i$ (where we
have denoted $G_i^L = \sum_{j=i}^{\infty} jc^L_j$, the corresponding
to $G_i$ for the solution $\{c^L_i\}$).

From now, to simplify the notation a bit, we will omit the $L$ in
both $c^L_j$ and $G^L_j$, as the full $c_j$ and $G_j$ will not be
mentioned anymore.  $W_{jk}$ will be used to denote $a_{jk}c^L_j
c^L_k - b_{jk} c^L_{j+k}$.

For any sequence $\{\Psi_j\}_{j \geq 1}$ it holds formally that:
\[
\frac{d}{dt} \sum_{j=1}^\infty \Psi_j c_j =
\frac{1}{2} \sum_{j,k=1}^\infty (\Psi_{j+k}-\Psi_j-\Psi_k) W_{jk}
\]

\begin{figure}[h]
  \center{\epsfig{figure=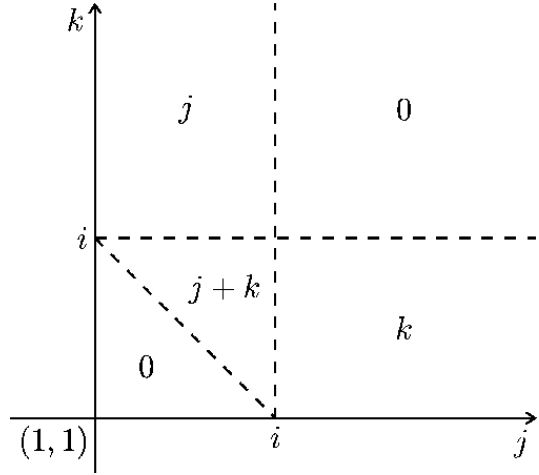, width=70mm}}
  \caption{Values of $\Psi_{j+k} - \Psi_j - \Psi_k$}
  \label{regions}
\end{figure}

In particular we can apply the previous relation
to $\Psi_j = j \cdot \one_{j \geq i}$ ($i \geq 1$) to get:
\begin{equation}
  \label{eq:dt0}
  \frac{d}{dt}\sum_{j=i}^\infty jc_j =
  \frac{1}{2} \sum_{j=1}^{i-1} \sum_{k=i-j}^{i-1} (j+k)W_{jk}
  + \sum_{j=1}^{i-1} \sum_{k=i}^\infty jW_{jk},
\end{equation}
and this equality is rigorously justified because the solution
$\{c_i\}$ has finite moments $\sum_{j=1}^\infty j^\mu c^L_j(t)$ of
every order $\mu \in \RR$ for every positive time $t$ (see lemma
\ref{lem:moments}), so the sums on both sides of the previous equality
converge uniformly and we can obtain the equation by means of standard
results on differentiation of uniformly convergent series of
functions. One way to obtain the expression on the right hand side is
to write the sum over $j, k$ as a sum over the regions depicted in
figure \ref{regions}, where the value of $\Psi_{j+k} - \Psi_j -
\Psi_k$ is indicated in each of them.

Due to hypothesis \ref{hyp:gBD}, $W_{jk} = 0$ if $\min\{j,k\} > N$.
Hence, for $i > 2N$ the first sum in (\ref{eq:dt0}) (which comprises
all pairs $j,k < i$ such that $j+k \geq i$) can be broken into those
terms where $j \leq N$ and those where $k \leq N$:
\begin{multline*}
  \frac{1}{2} \sum_{j=1}^{i-1} \sum_{k=i-j}^{i-1} (j+k)W_{jk}
  = \frac{1}{2} \sum_{j=1}^{N} \sum_{k=i-j}^{i-1} (j+k)W_{jk}
  + \frac{1}{2} \sum_{j=i-N}^{i-1} \sum_{k=i-j}^{i-1} (j+k)W_{jk}\\
  = \frac{1}{2} \sum_{j=1}^{N} \sum_{k=i-j}^{i-1} (j+k)W_{jk}
  + \frac{1}{2} \sum_{k=1}^{N} \sum_{j=i-k}^{i-1} (j+k)W_{jk}\\
  = \sum_{j=1}^{N} \sum_{k=i-j}^{i-1} (j+k)W_{jk},
\end{multline*}
where we have changed the order of the double sum and used the
symmetry of $(j+k) W_{jk}$. If $i > N$, the second sum in
(\ref{eq:dt0}) is nonzero only if $j \leq N$, so for $i > 2N$ we have
\begin{equation}\label{eq:dt}
  \frac{d}{dt}\sum_{j=i}^\infty jc_j =
  \sum_{j=1}^N \sum_{k=i-j}^{i-1} (j+k)W_{jk}
  + \sum_{j=1}^N \sum_{k=i}^\infty jW_{jk}.
\end{equation}
We rewrite the latter double sum:
\begin{multline}\label{eq:sum3}
  \sum_{j=1}^N \sum_{k=i}^\infty jW_{jk}
  = \sum_{j=1}^N \sum_{k=i}^{\infty} j (a_{jk} c_j c_k - b_{jk} c_{j+k}) \\
  = \sum_{j=1}^N \sum_{k=i-j}^{\infty} j a_{j,j+k} c_j c_{j+k}
  - \sum_{j=1}^N \sum_{k=i}^{\infty} j b_{jk} c_{j+k} \\
  = \sum_{j=1}^N \sum_{k=i-j}^{i-1} j a_{j,j+k} c_j c_{j+k}
  + \sum_{j=1}^N \sum_{k=i}^{\infty} j c_{j+k} (a_{j,j+k} c_j - b_{jk}) \\
  =: S_1 + S_2
\end{multline}
where we have denoted the two double sums as $S_1$, $S_2$
to mention them later.

We know that
\[
c_j(t) \leq Q_j z^j
\quad \text{for all } j=1,\ldots,N \text{ and } t \in [0,T].
\]
Hence, as $z < z_s$, we see that thanks to hypothesis \ref{hyp:db} and
for $j \in \set{1,\dots,N}$,
\begin{multline*}
  a_{j,j+k} c_j - b_{jk}
  \leq a_{j,j+k} Q_j z^j - a_{jk} Q_j \frac{Q_k}{Q_{j+k}}\\
  = Q_j a_{j,j+k}
  \paren{ z^j - \frac{a_{jk}}{a_{j,j+k}} \frac{Q_k}{Q_{j+k}} }.
\end{multline*}
Note that the term in parenthesis tends to $z^j - z_s^j$ as $k \to
\infty$ (thanks to hypotheses \ref{hyp:ajk_regular} and
\ref{hyp:Qjk}), so $S_2 \leq 0$ for $t \in [0,T]$ and $i$
sufficiently large. Then, continuing from \eqref{eq:dt}, using
\eqref{eq:sum3} and omitting $S_2$, we have for $i$ large that
\begin{equation}\label{eq:dt3}
  \frac{d}{dt}\sum_{j=i}^\infty jc_j
  \leq \sum_{j=1}^N \sum_{k=i-j}^{i-1} (j+k)W_{jk}
  + \sum_{j=1}^N \sum_{k=i-j}^{i-1} j a_{j,j+k} c_j c_{j+k}
\end{equation}

Using again $c_j \leq z_j := z^j Q_j$ for $j = 1,\dots,N$ and
\[
c_k=\frac{1}{k}(G_k-G_{k+1}) \quad \text{for all } k,
\]
rewrite \eqref{eq:dt3} as:
\begin{multline}\label{eq:dt3.5}
  \frac{d}{dt} G_i(t) \\
  \leq \sum_{j=1}^N \sum_{k=i-j}^{i-1}
  (j+k) \left( a_{jk}  c_j \frac{G_k - G_{k+1}}{k}
    - b_{jk}\frac{G_{j+k}-G_{j+k+1}}{j+k} \right) \\
  + \sum_{j=1}^N \sum_{k=i-j}^{i-1}
  j a_{j,j+k} c_j \frac{1}{j+k} (G_{j+k}-G_{j+k+1}) \\
  \leq  \sum_{j=1}^N \sum_{k=i-j}^{i-1}
  (j+k) \left( a_{jk}  z_j \frac{G_k - G_{k+1}}{k}
    - b_{jk}\frac{G_{j+k}-G_{j+k+1}}{j+k} \right) \\
  + \sum_{j=1}^N \sum_{k=i-j}^{i-1}
  j a_{j,j+k} z_j \frac{1}{j+k} (G_{j+k}-G_{j+k+1}) \\
  = \sum_{j=1}^N \sum_{k=i-j}^{i-1} \left(
    A_{jk}(G_k - G_{k+1}) - B_{jk}(G_{j+k} - G_{j+k+1})
  \right)
\end{multline}

Where
\begin{align*}
  & A_{jk} := \frac{(j+k) a_{jk} z_j}{k} \\
  & B_{jk} := \frac{(j+k)b_{jk} - j a_{j,j+k} z_j}{j+k}.
\end{align*}

Now we take any $\l < \bar{\l}< \frac{z_s}{z}$ (recall $\l$ appears
in the condition on $r_i$), and note that the following holds for
$k$ large enough:
\begin{equation}\label{eq:lambda}
  B_{jk} \geq \bar{\l}^j A_{jk} \quad \text{ for } j=1,\dots,N.
\end{equation}

The proof of this is easy, as (note that we can divide by $a_{jk}$ by
hypothesis \ref{hyp:gBD}):
\begin{multline}\label{A*B*}
  \frac{B_{jk}}{A_{jk}}
  = \frac{k b_{jk}}{(j+k) a_{jk} z_j}
  - \frac{j k a_{j,j+k} z_j}{(j+k)^2 a_{jk} z_j}\\
  = \frac{k a_{jk} Q_j Q_k}{(j+k) a_{jk} Q_{j+k} Q_j z^j}
  - \frac{j k a_{j,j+k}}{(j+k)^2 a_{jk}} \\
  = \frac{k}{j+k} \frac{Q_k}{Q_{j+k}} \frac{1}{z^j}
  - \frac{j k a_{j,j+k}}{(j+k)^2 a_{jk}},
\end{multline}
where the detailed balance hypothesis \ref{hyp:db} has been used to
pass to the second line. Now observe that the term with the negative sign
converges to 0 thanks to hypothesis \ref{hyp:ajk_regular}, and that
the other term
\begin{equation}
  \underset{k \to \infty}{\lim}
  \frac{k}{j+k} \frac{Q_k}{Q_{j+k}} \frac{1}{z^j}
  = \left( \frac{z_s}{z} \right)^j > \bar{\l}^j
\end{equation}
because of hypothesis \ref{hyp:Qjk}. Hence we have \eqref{eq:lambda}.

So, thanks to \eqref{eq:lambda}, we can continue from \eqref{eq:dt3.5}
and get, for $i$ large enough:
\begin{equation}\label{eq:dt4}
  \frac{d}{dt} G_i
  \leq \sum_{j=1}^N \sum_{k=i-j}^{i-1} A_{jk}
  \left(
    G_k - G_{k+1} - \bar{\l}^j(G_{j+k} - G_{j+k+1})
  \right)
\end{equation}
It is easy to see from hypothesis \ref{hyp:ajk_regular} that for
$j,m = 1,\dots,N$,
\[
\frac{A_{jk}}{A_{j,k+m}} \to 1 \quad \text{ as } k \to \infty.
\]
This means that for small variations of $k$, $A_{jk}$ changes little
when $k$ is large.  Take $\e$ such that
\begin{equation}\label{epsilon}
  \frac{1-\e}{1+\e} \geq \frac{\l}{\bar{\l}}.
\end{equation}
We can then find an $i_0 > 2N$ such that \eqref{eq:dt4} holds for $i \geq
i_0$ and we have, also for $i \geq i_0$:
\begin{equation}
  (1-\e)A_{j,i-j} \leq A_{jk} \leq (1+\e)A_{j,i-j}
  \quad \text{for } j=1,\dots,N, \quad k=i-j,\dots,i-1
\end{equation}
So for $i \geq i_0$ we can write from \eqref{eq:dt4}:
\begin{multline}\label{eq:dt5}
  \frac{d}{dt} G_i\\
  \leq \sum_{j=1}^N A_{j,i-j} \sum_{k=i-j}^{i-1}
  \left(
    (1+\e) (G_k - G_{k+1}) - (1-\e) \bar{\l}^j (G_{j+k} - G_{j+k+1})
  \right) \\
  = \sum_{j=1}^N A_{j,i-j}
  \left[
    (1+\e) (G_{i-j} - G_{i}) - (1-\e) \bar{\l}^j (G_{i} - G_{i+j})
  \right]
\end{multline}
From the hypothesis on $r_i$, for $j=1,\dots,N$ and $i > j$,
\begin{multline*}
  r_{i-j} - r_i = \sum_{k=1}^{j} (r_{i-k}-r_{i-k+1}) \\
  \leq \sum_{k=1}^{j} \l (r_{i-k+1}-r_{i-k+2})
  = \l (r_{i-j+1}-r_{i+1})
\end{multline*}
Apply this $j$ times to get:
\begin{equation}\label{eq:ri_final}
  r_{i-j} - r_i \leq \l^j (r_{i}-r_{i+j})
  \leq \bar{\l}^j \frac{1-\e}{1+\e} (r_{i}-r_{i+j}) \,,
\end{equation}
where we used \eqref{epsilon} together with $\l / \bar{\l} < 1$ to say
that
\[
\frac{1-\e}{1+\e} \geq \paren{\frac{\l}{\bar\l}}^j .
\]

If the sequence $\{r_i\}$ satisfies \eqref{eq:ri_final} then
$\{Cr_i\}$ also satisfies it, for any positive $C$.  Take $C>1$
sufficiently large so that
\begin{equation}\label{eq:M0}
  C r_i > M_0 \geq G_i(t) \quad \text{ for } i < i_0
  \, \text{and } t \leq T \,,
\end{equation}
where by $M_0$ we mean the density of the full initial data with no
truncation. Now define
\begin{align}
  M_i &:= G_i-Cr_i, \\
  H_i &:= (G_i-Cr_i)_+
\end{align}
We know $H_i(t) = 0$ for $i<i_0$ and $t<T$ because of \eqref{eq:M0}.

As the $Cr_i$ satisfy \eqref{eq:ri_final} we can write,
continuing from \eqref{eq:dt5}, for $i \geq i_0$:
\begin{equation}\label{eq:Mi}
  \frac{d}{dt} M_i
  \leq \sum_{j=1}^N A_{j,i-j}
  \left[
    (1+\e) (M_{i-j} - M_{i}) - (1-\e) \bar\l^j (M_{i} - M_{i+j})
  \right]
\end{equation}
Then, the same inequality holds for $H_i$: note that most of the
previous reorganization was done in order to have the term in $M_i$
as \emph{the only term with negative sign} in
\eqref{eq:Mi}. Otherwise we cannot justify writing the inequality in
terms of $H_i$ as is done next:
\begin{multline}\label{eq:Hi}
  \frac{d}{dt} H_i
  = \one_{M_i>0} \frac{d}{dt} M_i \leq \\
  \leq \one_{M_i>0} \sum_{j=1}^N A_{j,i-j}
  \left[
    (1+\e) (M_{i-j} - M_{i}) - (1-\e) \bar\l^j (M_{i} - M_{i+j})
  \right] \\
  \leq \sum_{j=1}^N A_{j,i-j}
  \left[
    (1+\e) (H_{i-j} - H_{i}) - (1-\e) \bar\l^j (H_{i} - H_{i+j})
  \right].
\end{multline}
(We have used $\one_{M_i > 0} M_i = H_i$ and $\one_{M_i > 0} M_k \leq
H_k$ for any $i,k$). Now we can sum this from $i=i_0$ to infinity
(note again that the sums are all convergent, as the solution $\{ c_j
\}$ with truncated initial data has finite moments of all orders) and
reorganize the terms:
\begin{multline}
  \frac{d}{dt} \sum_{i=i_0}^\infty H_i
  \leq \sum_{i=i_0}^\infty \sum_{j=1}^N A_{j,i-j}
  \left[
    (1+\e) (H_{i-j} - H_{i}) - (1-\e) \bar\l^j (H_{i} - H_{i+j})
  \right] \\
  = \sum_{j=1}^N \sum_{i=i_0-j}^\infty H_i
  \paren{ (1+\e) A_{ji} - (1-\epsilon) \bar{\lambda}^j A_{j,i-j}}
  - \sum_{j=1}^N \sum_{i=i_0}^\infty H_i (1+\epsilon) A_{j,i-j}\\
  + \sum_{j=1}^N \sum_{i=i_0+j}^\infty H_i (1-\epsilon) \bar\l^j A_{j,i-2j}\\
  = \sum_{j=1}^N \sum_{i=i_0-j}^{i_0-1} A_{ji} (1+\e) H_i
  + \sum_{i=i_0}^\infty H_i  (1+\e) \sum_{j=1}^N
  \left[ A_{ji} - A_{j,i-j} \right]\\
  + \sum_{i=i_0}^\infty H_i (1-\e) \sum_{j=1}^N
  \bar\l^j \left[ A_{j,i-2j} - A_{j,i-j} \right]
  - \sum_{j=1}^N \sum_{i=i_0}^{i_0+j-1} H_i A_{j,i-2j} \bar\l^j (1-\e) \\
  =: T_1 + T_2 + T_3 + T_4,
\end{multline}
where the $T_i$ (i=1,2,3,4) are the sums above. Observe that $T_4$ is
negative and $T_1$ only contains terms in $H_i$ for $i < i_0$, so it
is directly zero (recall \eqref{eq:M0}). Also, note that for
$j=1,\dots,N$ and $i \geq i_0$ we have, by using hypothesis
\ref{hyp:ajk_regular2}, that
\begin{multline*}
  \abs{A_{ji} - A_{j,i-j}}
  = \abs{
    \frac{j+i}{i} a_{ji} z_j - \frac{i}{i-j} a_{j,i-j} z_j
  }\\
  \leq z_j a_{ji} \abs{ \frac{j+i}{i} - \frac{i}{i-j} }
  + z_j \frac{i}{i-j} \abs{ a_{ji} - a_{j,i-j} } \\
  \leq z_j K (i+j) \frac{j^2}{i(i-j)}
  + z_j \frac{i}{i-j} K_a,
\end{multline*}
which is easily seen to be bounded by a certain constant $A'$ for
$j=1,\dots,N$ and $i > N$. Hence, the coefficient of $H_i$ in
$T_2$ and $T_3$ is bounded by a certain constant $A$ independent of
$j$ and $i$ and then
\begin{equation*}
  \frac{d}{dt} \sum_{i=i_0}^\infty H_i
  \leq A \sum_{i=i_0}^\infty H_i.
\end{equation*}
Gronwall's lemma then shows that $H_i(t)=0$ for $i \geq i_0$ and $t
\in [0,T]$; that is to say $G_i(t) \leq C r_i$. This proves our claim.

\subsection{Proof of the main theorem}
\label{sec:Other_Results}

Finally, we arrive at the proof of theorems \ref{thm:main} and
\ref{thm:final}, which is not difficult once the proposition of the
previous section has been established.

\begin{proof}[Proof of Theorem \ref{thm:main}]
  Let $c$ be an admissible solution that converges weak-$\ast$ in $X$
  to an equilibrium of mass $\rho < \rho_s$, which must be given by
  $\{Q_j \bar{z}^j \}_{j\geq 1}$ for some $0 \leq \bar{z} < z_s$ (see theorem
  \ref{thm:equilibria}). We will prove that the orbit of any such
  solution must be relatively compact in $X$ and hence the convergence
  must be strong.

  Pick $z \in ]\bar{z},z_s[$. As we know that $c_j \to Q_j \bar{z}^j$
  when $t\to\infty$ for all $j$, we can find a $t_0>0$ so that
  \[
  c_j(t) \leq Q_j z^j
  \quad \text{for all } j=1,\ldots,N \text{ and } t\geq t_0.
  \]
  As Lemma \ref{lem:findrk} ensures, we can always find a sequence
  $\{r_i\}$ tending to zero as $i \to \infty$ that satisfies the
  conditions in Proposition \ref{prp:G} with $G_i(t_0)$ instead of
  $G_i(0)$. We can apply the proposition, with the $z$ we have chosen,
  to $\{c_j(t+t_0)\}_{j \geq 1}$ (which is a translation in time of
  the solution $c$ and thus is a solution itself) and deduce that for
  some $C>0$, $k_0 \geq 1$ and all $t>t_0$,
  \[
    G_i(t) \leq C r_i \quad \text{ for } i \geq k_0.
  \]
  As $\{r_i\}$ tends to zero, this bound says that the solution $c$ is
  relatively compact in $X_+$, and we have finished.
\end{proof}

\begin{proof}[Proof of Theorem \ref{thm:final}]
  Suppose that $c$ is an admissible solution of (\ref{eq:gBD}) in
  $[0,+\infty[$ with initial data $c(0) = c^0 \in X_+$. Theorem
  \ref{thm:weak-star-convergence} shows that $c$ converges weak-$\ast$
  in $X$ to an equilibrium of mass $\rho$ for some $0 \leq \rho \leq
  \rho_0$.

  If $\rho_0 < \rho_s$, then this convergence is also strong (by
  theorem \ref{thm:main}) and hence $\rho=\rho_0$.

  If $\rho_0 = \rho_s$ (with $\rho_s < \infty$), then $\rho \leq
  \rho_s$. But if $\rho < \rho_s$ then again the convergence must be
  strong and $\rho=\rho_s$, which is a contradiction. Hence,
  $\rho=\rho_0=\rho_s$ and we see the convergence is strong because of
  lemma \ref{lem:weak-strong-convergence}.

  Finally, if $\rho_0 > \rho_s$, then $\rho \leq \rho_s$ and it must
  be $\rho = \rho_s$ or otherwise the convergence is strong, which is
  not possible given that $\rho_0 > \rho$.
\end{proof}


\end{document}